\documentclass[12pt]{iopart}
\usepackage{iopams}  
\expandafter\let\csname equation*\endcsname\relax
\expandafter\let\csname endequation*\endcsname\relax

\usepackage{cite}
\usepackage{graphicx} 
\usepackage{dcolumn}
\usepackage{bm}
\usepackage{mathtools}
\usepackage{hyperref}
\usepackage{graphicx}
\usepackage{amsmath}
\usepackage{braket}
\usepackage{amsfonts}
\usepackage{amssymb}
\usepackage{bm}
\usepackage[usenames,dvipsnames]{color}
\usepackage{float}
\usepackage{amsmath,amssymb}
\usepackage{amsfonts}
\usepackage{csquotes}
\usepackage{comment}
\usepackage{bbold}
\usepackage{romannum}
\usepackage{mathrsfs}
\hypersetup{
	colorlinks=true,
	citecolor=blue,
	linkcolor=blue,
	urlcolor=blue}

\usepackage{subfig}
\usepackage{tikz}
\usepackage{pgffor} 
\usepackage{hyperref}

\begin{document}

\title[]{Flat Bands in Three-dimensional Lattice Models with Non-trivial Hopf Index}

\author{Ivan Dutta$^1$$^,$$^2$$^,$$^*$,Kush Saha$^1$$^,$$^2$}

\address{$^1$ National Institute of Science Education and Research, Jatni, Odisha 752050, India}
\address{$^2$ Homi Bhabha National Institute, Training School Complex, Anushakti Nagar, Mumbai 400094, India}
\address{$^*$ Author to whom any correspondence should be addressed.}

\eads{\mailto{ivan.dutta@niser.ac.in},   \mailto{kush.saha@niser.ac.in}}

\begin{abstract}
We report the presence of exactly and nearly flat bands with non-trivial topology in three-dimensional (3D) lattice models. We first show that an exactly flat band can be realized in a 3D lattice model characterised by a 3D topological invariant, namely Hopf invariant. In contrast, we find another distinct 3D model,  exhibiting both 2D Chern and 3D Hopf invariant, namely Hopf-Chern insulator, that can host nearly or perfect flat bands across different 2D planes. Such a Hopf-Chern model can be constructed by introducing specific hopping along the orthogonal direction of a simple two-orbital 2D Chern insulator in the presence of in-plane nearest-neighbor and next-nearest hopping among different orbitals. While the Chern planes host nearly perfect flat bands, the orthogonal planes can host both perfect or nearly perfect flat bands with zero Chern number at some special parameter values.  Interestingly, such a 3D lattice construction from 2D allows finite Hopf invariant too. Finally, we show that higher Chern models can also be constructed in the same lattice setup with only nearest and next-nearest hopping, but the appearance of flat bands along high-symmetric path in the Brillouin zone requires longer-range hopping. We close with a discussion on possible experimental platforms to realize the models.
\end{abstract}


\maketitle

\section{Introduction}
The discovery of topological insulators has culminated in a new classification scheme for free fermion theories, namely {\it tenfold} way for solids that relies on the interplay between symmetries and dimensionality \cite{PhysRevB.78.195125,Kitaev_2009,PhysRevB.76.045302}. While the tenfold symmetry classification predicts no topological phases in three dimensions (3D), particularly in the absence of time-reversal, particle-hole and sub-lattice symmetries (class A), there exists a unique class of 3D magnetic insulators beyond {\it tenfold} table with gapless surface mode in presence of exactly one valence and one conduction band. These insulators are termed as Hopf insulators and they are characterised by an integer-valued ($\mathbb{Z}$) invariant, namely linking number. The first ever genuinely three dimensional Hopf insulator was proposed a decade back \cite{PhysRevLett.101.186805} and subsequently there have been several proposals and studies on Hopf insulators \cite{PhysRevB.88.201105,PhysRevB.94.035137,PhysRevB.95.161116,PhysRevB.103.045107,PhysRevB.103.014417, PhysRevLett.118.147003, PhysRevB.105.045139,PhysRevResearch.3.033045,PhysRevB.106.075124,Yuan_2017,PhysRevLett.130.057201,PhysRevLett.119.156401,PhysRevB.101.155131,PhysRevB.97.155140,PhysRevA.99.043619,deng2018probe,PhysRevB.102.035101,PhysRevLett.123.266803,PhysRevB.96.041202,Wang:2023dsg,leng2022n,cook2022multiplicative,PhysRevResearch.1.022003,PhysRevB.107.045130,Graf_2022,PhysRevLett.125.053601,PhysRevLett.130.057201,PhysRevB.108.125101}.

Recently, there has been tremendous research interest in studying topology together with dispersionless bands. This is because the dispersionless bands can host a plethora of interesting physical phenomena such as inverse Anderson insulators\cite{PhysRevLett.96.126401,PhysRevB.82.104209}, multifractality at weak disorder\cite{PhysRevB.87.125428,doi:10.1143/JPSJ.72.2015}, Hall ferromagnetism\cite{PhysRevB.65.081307,PhysRevLett.100.136404}, chiral spin liquid\cite{PhysRevB.99.174418, PhysRevLett.114.037203}, Sachdev-Ye-Kitaev (SYK)\cite{PhysRevB.108.064202} physics and many more\cite{PhysRevB.88.224203,Flach_2014,PhysRevLett.113.236403,PhysRevB.91.235134,PhysRevB.96.161104,PhysRevLett.116.245301,PhysRevLett.114.245503,PhysRevLett.114.245504,PhysRevA.82.041402,PhysRevLett.99.070401,PhysRevB.104.085144,SciPostPhys.15.4.139,23746149.2018.1473052,Kuno_2020,Chen_2019,Masumoto_2012,Rhim2020-jy,Oh2022-bv}. Additionally, mixing flat bands with topology makes them promising venues for studying lattice version of fractional Chern insulating (FCI) states at various fillings\cite{PhysRevLett.106.236803,PhysRevLett.106.236802,PhysRevLett.106.236804}, which has led to a great volume of works focusing on flat bands with non-trivial topology, particularly in two dimensions (2D)\cite{PhysRevB.86.241112,PhysRevA.83.063601,PhysRevB.86.085311,Bergman,PhysRevB.97.195101,PhysRevB.104.195128,PhysRevA.91.033604, sur2021insulatormetal}.

Despite significant progress, realizing a stable Fractional Chern Insulator (FCI) state without any external magnetic field is very limited due to the lack of suitable experimental platforms. Although many experimental platforms use an external magnetic field\cite{science.aan8458,Xie2021-zs}, FCIs without the external magnetic field has only been observed recently in twisted rhombohedral-stacked transition-metal-dichalcogenide bilayers (MoTe$_2$)  \cite{Cai2023-rr,nature:v:622}. This limitation is attributed to the intrinsic flatness of the band structure within the lattice, which is one of the crucial factors for mimicking Landau levels and determining the stability of FCIs. It is commonly believed: the more the bands are flat, more the FCI states are stable. However,  perfect flat bands with finite Chern number are restricted by the following proposition: \enquote{exactly flat band, non-zero Chern number, and local hopping} cannot all be simultaneously realized\cite{Chen_2014, sathe2023topological}. Additionally, while FCI states are mostly studied in 2D, a few recent series of experimental evidences of 3D quantum  Hall effect in 3D materials \cite{Li2021,Tang201912,PhysRevB.101.161201} has led to the theoretical proposal for 3D fractional quantum Hall effect\cite{PhysRevLett.124.096603,PhysRevB.101.235168,PhysRevB.85.041104}. Moreover, the physics of flat band with non-trivial topology in 3D \cite{PhysRevB.88.205107,PhysRevB.85.041104, huang2023three, PhysRevLett.111.226403, annurev-conmatphys-031214-014749, PhysRevX.11.031017} is limited. Furthermore, many of the interesting phenomena are observed if the flat bands are isolated from the rest of the bands as interactions are not bound to mix states from adjacent bands \cite{Hwang2021-bg}. In view of the above, we aim to answer to the following questions: is it possible to construct a three dimensional lattice model exhibiting {\it exact} flat bands with finite topological indices but restricting only to short-ranged hopping? Can such a lattice model host isolated flat bands?


Affirmatively addressing these questions, we present two distinct constructions of three-dimensional lattice models wherein the band exhibits exact flatness along with non-trivial topology. 
The first model, characterised by a 3D Hopf invariant, can host  a perfectly flat isolated band across all three two-dimensional planes. In contrast, the second model, characterised by both Hopf and Chern invariant, namely Hopf-Chern insulator, accommodates a nearly flat band within a specific two-dimensional plane where the Chern number is finite. Notably, the pure Hopf model incorporates hopping terms up to next-to-next-nearest neighbour (NNNN) and the appearance of flat bands does not require any additional range of hopping beyond NNNN. In contrast, the Hopf-Chern model restricted to next-nearest-neighbour (NNN) hopping gives rise to nearly flat bands. Throughout this manuscript we mostly focus on the Hopf-Chern case as it is comparatively closer to realize in experiments. We show that such a construction of Hopf-Chern 3D lattice can be obtained from a 2D Chern insulator with specific hopping along the third direction. We further chart out phase diagram of this 3D  model Hamiltonian for different configurations of parameters. We find that both the Hopf and Chern phases appear together and symmetrically with respect to trivial insulating phases. We further show that higher Chern phases can also be constructed out of the same lattice model for a particular choice of parameters, restricting to only NN and NNN hoppings. However, such a construction does not allow flat bands in all high-symmetric path in the Brillouin zone (BZ). To this end, we comment on the feasibility of realizing such Hopf models with trivial and non-trivial Chern numbers in experiments.

\begin{figure}
\centering
\includegraphics[width=0.66\linewidth]{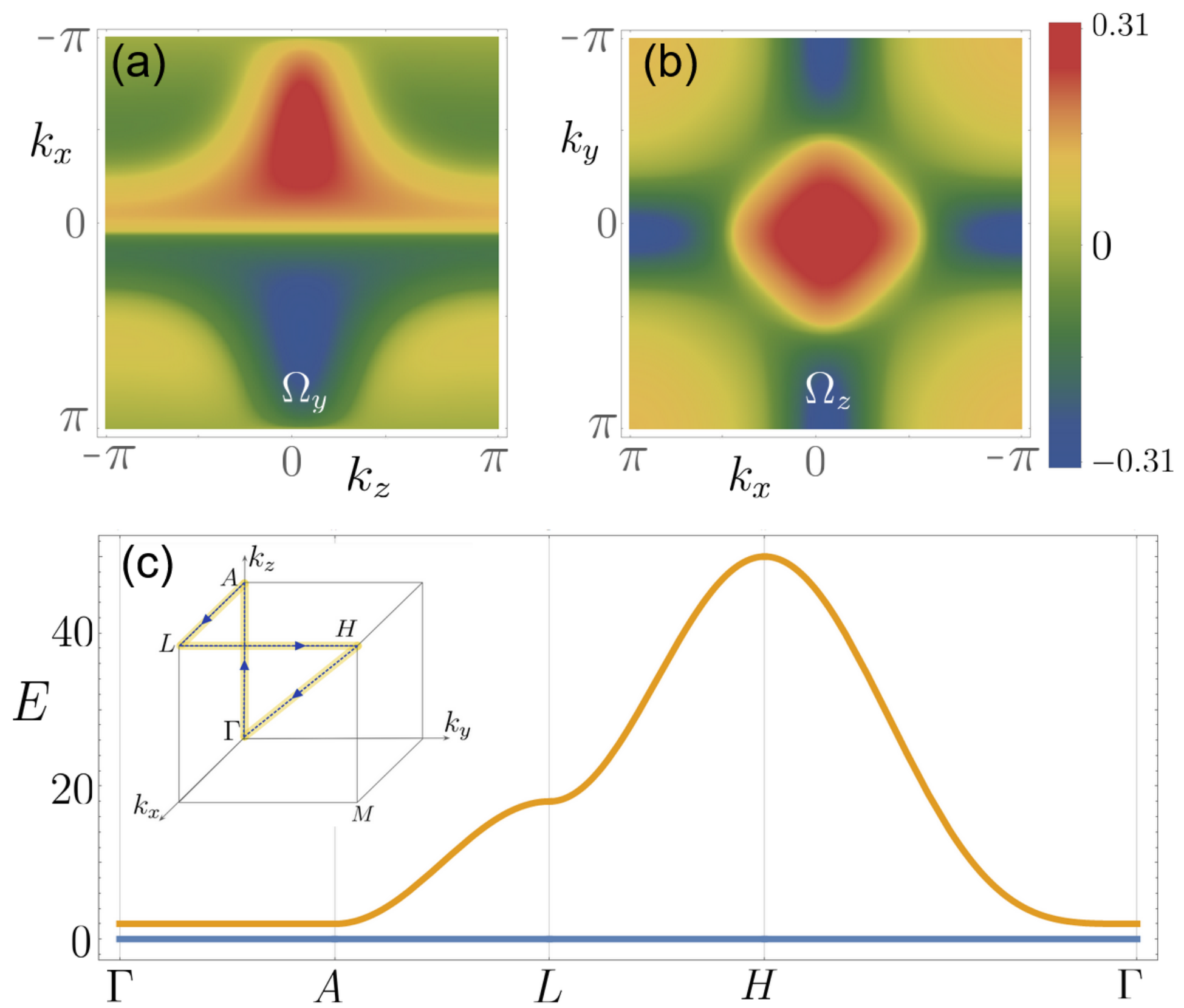}
	\caption{Finite Berry curvature $\Omega$ along (a) $k_y = 0$ and (b) $k_z = 0$ plane. (c) Energy spectrum of Eq.~(\ref{eq:hopf}) in the presence of $d_0 ({\bf k})$ term along the high-symmetry path revealing a perfectly flat isolated valence band for $h = -2$.}
	\label{Hopf_flat_bands}
\end{figure}

\section{Exactly Flat Hopf band in 3D} We begin with a two-band model on a cubic lattice in which each site has two orbital states $|A\rangle$, $|B\rangle$. The tight binding Hamiltonian can be written in compact form as,
\begin{align}
H_{\textrm {hopf}} = \sum_{\Vec{\mathbf{r}}, \mu, \nu} \ket{\Vec{\mathbf{r}} + \Vec{e_{\mu}}}\:\:\Gamma^{\mu\nu}\:\:\bra{\Vec{\mathbf{r}} + \Vec{e_{\nu}}}
\label{eq:Hopf_TB}
\end{align}
where $\mu$,$\nu$ are characterised by their position in the array of hopping direction $(0,x,y,z,xy,yz,zx,2x,2y,2z)$. $\Vec{e}_{\mu/\nu}$ is the position vector of the hopping and $\Gamma$'s are matrices of hopping strength between two orbitals along this 10 different directions. All the $\Gamma$'s are illustrated in Appendix \ref{Hopf Flat}. In momentum space, this reads off 
\begin{align}
H_{\mbox{hopf}} ={\bf d({\bf k})}\cdot {\bf\sigma},
\label{eq:hopf}
\end{align} where 
\begin{align}
&d_x(\mathbf{k})= 2 (N_1 N_2 + N_3 N_4)\nonumber\\ 
&d_y(\mathbf{k}) = 2 (N_2 N_3 - N_1 N_4)\nonumber\\
&d_z(\mathbf{k}) = \sum_{i=1}^4\, (-1)^{i+1} N_i^2.
\end{align}
Here $\sigma$ denotes the Pauli matrices, $N_1 = \sin(k_x)$, $N_2 = \sin(k_z)$, $N_3 = \sin(k_y)$, $N_4 = \sum_{i = x,y,z}\cos(k_i) + h$, $h$ being the parameter of the Hamiltonian. For $-3 < h < 3$, this Hamiltonian in Eq.~(\ref{eq:Hopf_TB}) hosts gapped bulk but gapless surface modes protected by Hopf invariant, denoted as $(n_0\in \mathbb{Z}; 0,0,0)$ \cite{PhysRevLett.101.186805, PhysRevB.88.201105}. The $(0,0,0)$ form represents trivial Chern insulating phases for all three planes of the 3D lattice. The reason for zero Chern number ($C=0$) in all 2D planes is attributed to the fact that the Berry curvature is symmetric as shown in Fig.~(\ref{Hopf_flat_bands}a-b). We note that, for $C\ne 0$ in any 2D planes, the Hopf invariant belongs to finite group ${\mathbb Z}_{2. {\textrm GCD}(C_x,C_y,C_z)}$, where $C_x$, $C_y$ and $C_z$ represent the Chern numbers in y-z, z-x, x-y planes respectively and GCD denotes greatest common divisor \cite{PhysRevB.94.035137}. For example, if a Hopf-Chern model has Chern numbers (2,0,0), the GCD is 2. Consequently, the 3D invariant will be $\mathbb{Z}_4$, indicating that only four distinct 3D topological phases are possible in the system.

To find flat bands, we introduce $d_0 ({\bf k}) = \sum_{i=1}^4\, N_i^2 $, which does not involve any additional range of hopping beyond the hopping associated with $d_x,d_y,d_z$. Interestingly, this leads to a perfect flat band with $E=0$ hosting compact localized state as illustrated in Appendix \ref{Hopf-CLS}. The corresponding band spectrum in the full 3D BZ is shown in Fig.~(\ref{Hopf_flat_bands}c). This confirms that, unlike the Chern number, the Hopf invariant does not impose any topological obstruction for flattening the band spectrum.
\section{Hopf-Chern Flat Bands}
\subsection{Nearly Flat Chern band in 2D} 
\begin{figure}
\centering
	\includegraphics[width=0.69\linewidth]{{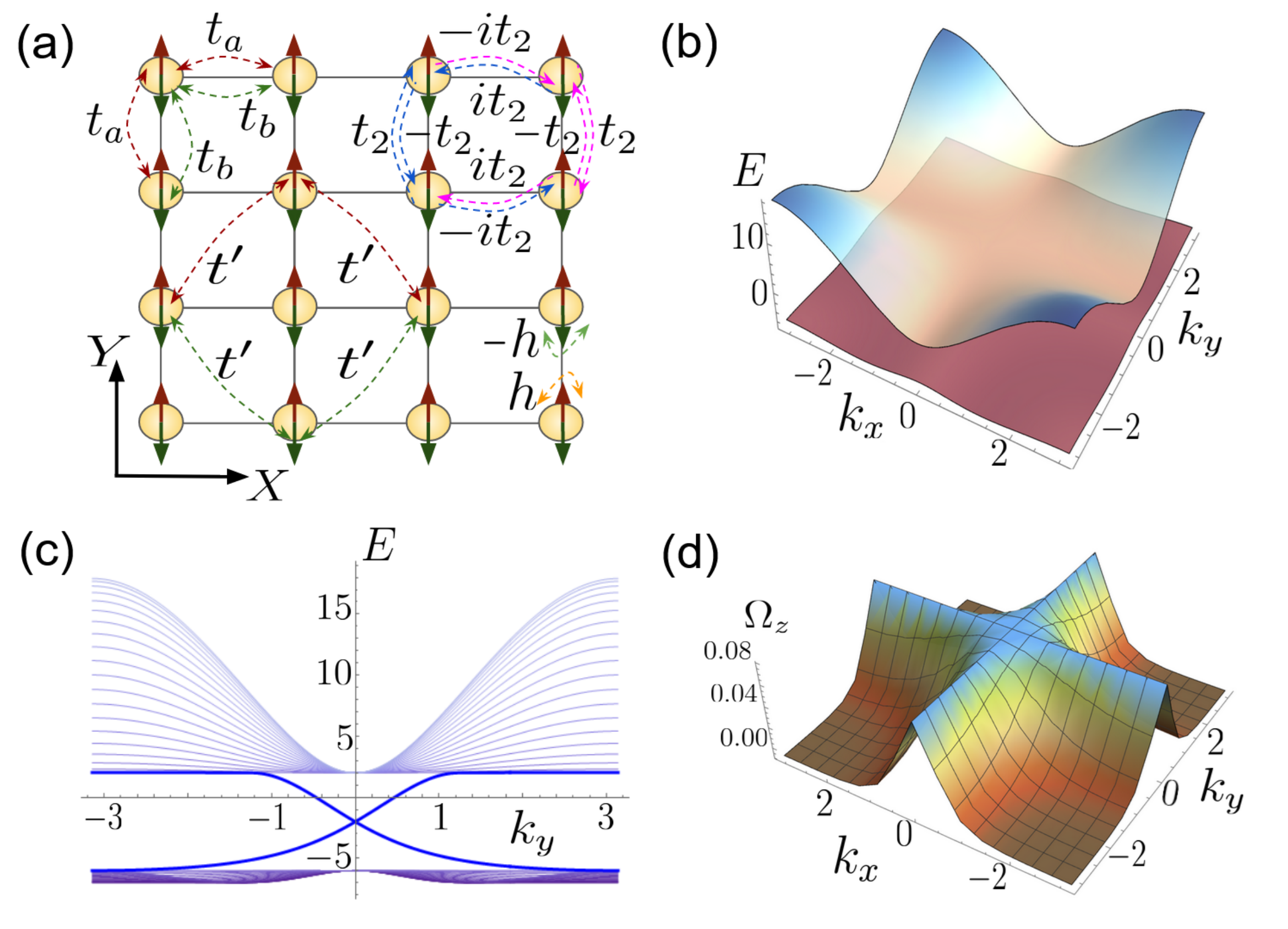}}
	\caption{(a) The schematics of two-orbital 2D tight binding model where the A and B orbitals are represented by up-arrow and down-arrow, respectively. Distinct colored lines such as brown, green, magenta and blue denote the hopping between A to A, B to B, A to B and B to A respectively.  The parameter $t_a (t_b)$ represents NN hopping strength between A (B) orbital states, while $t'$ denotes NNN strength between similar orbital states. The hopping magnitude between distinct orbitals along the x and y direction is represented by $t_2$ and they differ by a phase $\pm \pi/2$. (b) The bulk band spectra of Eq.~(\ref{eq:2D_Bulk_Ham}) with a nearly flat band for the parameter $ h=-4$, $t_2 = -2 $, $t' = 0.5$ ,  $t_a=1$ and $t_b=-3$. (c) Gapless edge spectrum in the $E-k_y$ plane and (d) Plot of Berry curvature of the nearly flat band.}
	\label{fig:2D_Model_Plot}
\end{figure}

In contrast to the pure Hopf insulators, the Hopf-Chern models can be constructed systematically starting from a 2D Chern insulator. Further it allows us to construct Hopf-Chern model with higher Chern number as will be evident shortly. Thus here we take an independent route. We consider a two-band model on a 2D square lattice in which each lattice site contains two orbital states $|A\rangle$, $|B\rangle$ as before. The tight-binding Hamiltonian reads off

\begin{align}
H_{\textrm 2D} &=  \sum_{i\sigma}\epsilon_{\sigma}c_{i\sigma}^{\dagger}c_{i\sigma}+ \sideset{} {_{x} }\sum_{<ij>,\sigma,\sigma'}t_{\sigma\sigma'}^x c_{i\sigma}^{\dagger} c_{j\sigma'} + \sideset{} {_{y} }\sum_{<ij>\sigma,\sigma'}t_{\sigma\sigma'}^y c_{i\sigma}^{\dagger} c_{j\sigma'} +\sum_{\ll ij\gg\sigma,\sigma'}t'_{\sigma\sigma'} c_{i\sigma}^{\dagger} c_{j\sigma'}
\label{eq:Ham0}
\end{align}
where $\sum_x$ and $\sum_y$ terms imply the summation for electrons hopping along the x and y directions respectively, the angular bracket $<>$ and $\ll\gg$ represent NN and NNN lattice sites, respectively; $t^{x/y},t'$ are corresponding hopping strengths, $\epsilon_{\sigma}$ denotes onsite energy and $\sigma\in (A,B)$. In Fig.~(\ref{fig:2D_Model_Plot}a), the hopping between different sites and species are shown. The parameters $t_{\sigma\sigma'}^{\alpha}$ and $t'_{\sigma\sigma'}$ have the form
\begin{align}
t_{\sigma\sigma'}^{\alpha}=\left(\begin{matrix}t_{AA}^{\alpha}& t_{AB}^{\alpha}\cr t_{BA}^{\alpha}&t_{BB}^{\alpha} \end{matrix}\right),   t'_{\sigma\sigma'}=\left(\begin{matrix}t'_{AA}& 0\cr 0&t'_{BB} \end{matrix}\right)
\label{eq:hoppings}
\end{align}
where $\alpha \in (x,y)$. Assuming $t_{AA}^x = t_{AA}^y =t_a, t_{BB}^x = t_{BB}^y = t_b$, $t_{AB}^x = t_{BA}^x =-i t_2$, $t_{AB}^y = -t_{BA}^y =- t_2$, $\epsilon_{A}=-\epsilon_{B}=h$, and $t'_{AA}=t'_{BB}=t'$,  the Hamiltonian in Eq.~(\ref{eq:Ham0}) can be expressed in momentum space as 
\begin{align}
H_{\textrm 2D}(\textbf{k})=d_0 ({\bf k}) \mathbb{1}+{\bf d({\bf k})}\cdot {\bf\sigma},
\label{eq:2D_Bulk_Ham}
\end{align}
where $d_0=4t'\cos k_x\cos k_y + (t_a + t_b) (\cos k_x+\cos k_y)$, ${\bf d(\bf k)}=(2t_2\sin k_x,2t_2\sin k_y, (h+(t_a-t_b)(\cos k_x+\cos k_y))$; ${\bf \sigma}=(\sigma_x,\sigma_y,\sigma_z)$ are the Pauli matrices, ${\bf k}=(k_x,k_y)$ is the crystal momentum. The energy dispersion of Eq.~(\ref{eq:2D_Bulk_Ham}) is $E({\bf k})=d_0({\bf k})\pm \sqrt{|{\bf d(k)}|^2}$. It is apparent that one of the bands may become nearly flat if the $k$ dependence of $d_0({\bf k})$ approximately cancels the $k$  dependence of $\sqrt{|{\bf d(k)}|^2}$. Indeed, this scenario is achievable if we take $(t_a-t_b)>0$, $(t_a-t_b)=-h=-2t_2$, and $(t_a+t_b)=-4t'$. This leads one of the bands nearly flat and the flatness is nearly $1/8$ of the band gap.

The corresponding plot is shown in Fig. (\ref{fig:2D_Model_Plot}b). It is perhaps possible to enhance the flatness by utilizing a methodical numerical minimization approach. It is worth noting that both the NN ($t_a, t_b$) and NNN ($t'$) hopping are required to obtain nearly flat bands in the current setting. We note that the present setup does not have any finite flux through each plaquette, but hopping along particular directions have some gauge fluxes. This differs from earlier studies where flat bands with non-trivial topology on a square lattice were mainly obtained in the presence of a specific finite flux \cite{PhysRevLett.106.236802,PhysRevLett.106.236804} in a fine-tuned parameter regime. 

To find the topology of the two-band model in Eq.~(\ref{eq:2D_Bulk_Ham}), we compute Berry curvature (cf. Fig.~\ref{fig:2D_Model_Plot}d) ($\Omega({\bf k})$) of the occupied band ($\epsilon_F=0$) of bulk band spectrum. Subsequently, we compute Chern number  $C=\frac{1}{2\pi}\int d{\bf k}\,\Omega ({\bf k})$ and it turns out the model in Eq.~(\ref{eq:2D_Bulk_Ham}) exhibits three distinct Chern numbers. For $0<h<2 (t_a-t_b)$, we obtain $C=-1$, whereas for $-2 (t_a-t_b)<h<0$, $C=1$, otherwise we obtain zero Chern number. This is manifested in the edge spectrum illustrated in Fig.~(\ref{fig:2D_Model_Plot}c). It is evident that $d_0$ does not contribute to the topological invariant, however the parameters $t',t_a,t_b$ in $d_0$ play important role in making the Chern band nearly flat as discussed in the preceding section.

\begin{figure}
\centering
	\includegraphics[width=0.6\linewidth]{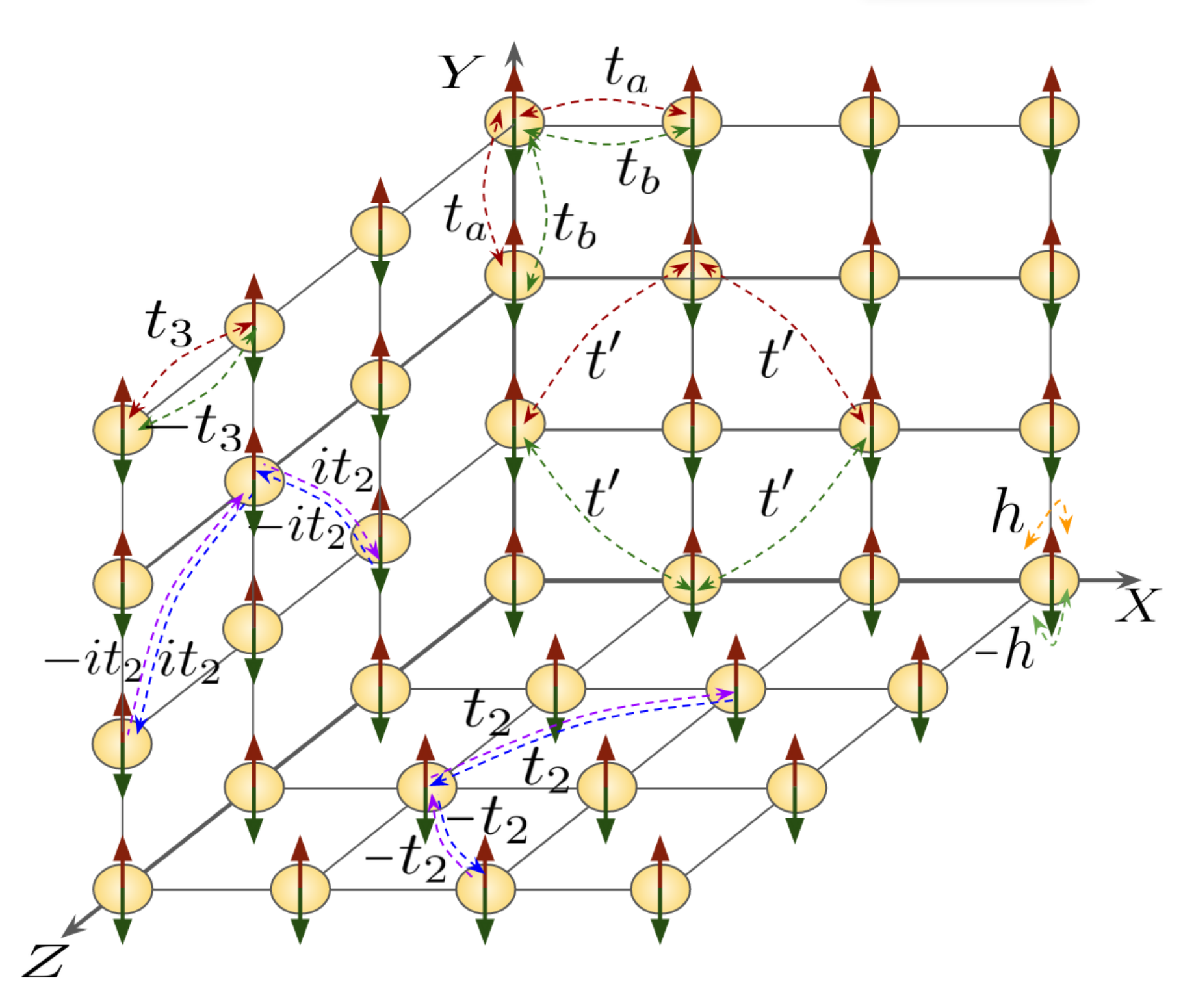}
	\caption{The schematics of 3D tight binding model. Brown, green, purple and blue colored arrows represent hopping between A to A, B to B, A to B and B to A orbitals, respectively. Note that the $x-y$ planes allow NN and NNN hopping between similar orbitals. In contrast, the $y-z$ and $x-z$ planes introduce NNN hopping between opposite orbital states. This is a minimal lattice model that can give rise to Hopf number in contrast to the standard Hopf insulator where more terms are required to be incorporated\cite{PhysRevLett.101.186805}.}
	\label{fig:3D_Lattice_Hopf}
\end{figure}

\subsection{Construction of non-trivial flat bands in 3D}
To construct a three-dimensional insulator with non-trivial flat bands, we can simply stack several layers of the 2D Chern insulators with weak nearest-neighbor interaction ($t_3$). This adds a term $2\,t_3 \cos k_z$ with $\sigma_z$ in Eq.~(\ref{eq:2D_Bulk_Ham}). As $t_3$ increases, the flatness gradually diminishes although the bulk spectrum maintains a gap with nontrivial Chern number in $k_x-k_y $ plane. At $t_3/h = \pm\frac{1}{2}$, the bulk gap closes and a trivial metallic phase is obtained. Indeed, this is a {\it trivial} construction of 3D insulators with flat bands exhibiting only Chern invariant.  

In contrast, we now provide a genuinely 3D lattice model which exhibits flat bands with both Chern and 3D Hopf invariants. To construct such a 3D lattice, we allow non-identical orbital hopping between next-nearest-neighbor in two different planes in contrast to the 2D case discussed earlier. This lattice design is derived as a simplified version of the 3D Hopf Model, achieved by selectively deactivating specific hopping terms and tuning other hopping parameters. The schematic of such construction is illustrated in Fig.~(\ref{fig:3D_Lattice_Hopf}). Note that we retain NN and NNN hopping between similar orbitals with similar strength in the $x-y$ plane as in 2D case. This leaves $d_0({\bf k})$ invariant. In the planes $x-z$ and $y-z$, the orbital-flipping NN terms of the previous 2D lattice act as NNN hopping.    

The momentum space Hamiltonian for such a 3D lattice construction reads off
\begin{align}
H_{\textrm 3D}(\textbf{k}) = d_0(\textbf{k})\mathbb{1} + d_x(\textbf{k})\sigma_x + d_y(\textbf{k})\sigma_y + d_z(\textbf{k})\sigma_z,
\label{eq:3DMomentum_Ham}
\end{align}
where $k=(k_x,k_y,k_z)$ is a 3D crystal momentum; $d_0(\textbf{k})$ is same as before, however $d_i(\textbf{k})$s  ($i\in\{x,y,z\}$) are modified as  $d_x(\textbf{k}) = 2 t_2 (\sin k_x \sin k_z +\cos k_z \sin k_y)$, $d_y(\textbf{k}) = 2 t_2 (-\sin k_y\sin k_z +\cos k_z \sin k_x)$, $d_z(\textbf{k})=(h+(t_a-t_b)(\cos k_x+\cos k_y))+2t_3\cos k_z$, due to the presence of hopping matrix elements along the $k_z$ direction. We note that the form of the 3D lattice model Hamiltonian has similar momentum structure (except a few other terms in the current model) with the rotated 2D Chern insulator introduced by Kennedy, where rotation is along the $z$ direction and the third momentum  $k_z$ plays the role of angle of rotation. The model from \cite{PhysRevB.94.035137} can be recovered from Eq.~(\ref{eq:3DMomentum_Ham}) at the special parameter values  $t_1=t_2=t_a= - t_b = 1/2$, $t_3=0$ and $t’=0$.  It is important to note that such a construction of 3D  insulators by rotation may not be a generic construction, particularly for arbitrary Chern phases and it requires a detailed investigation in its own right. 

\begin{figure}
\centering
\includegraphics[width=0.85\linewidth]{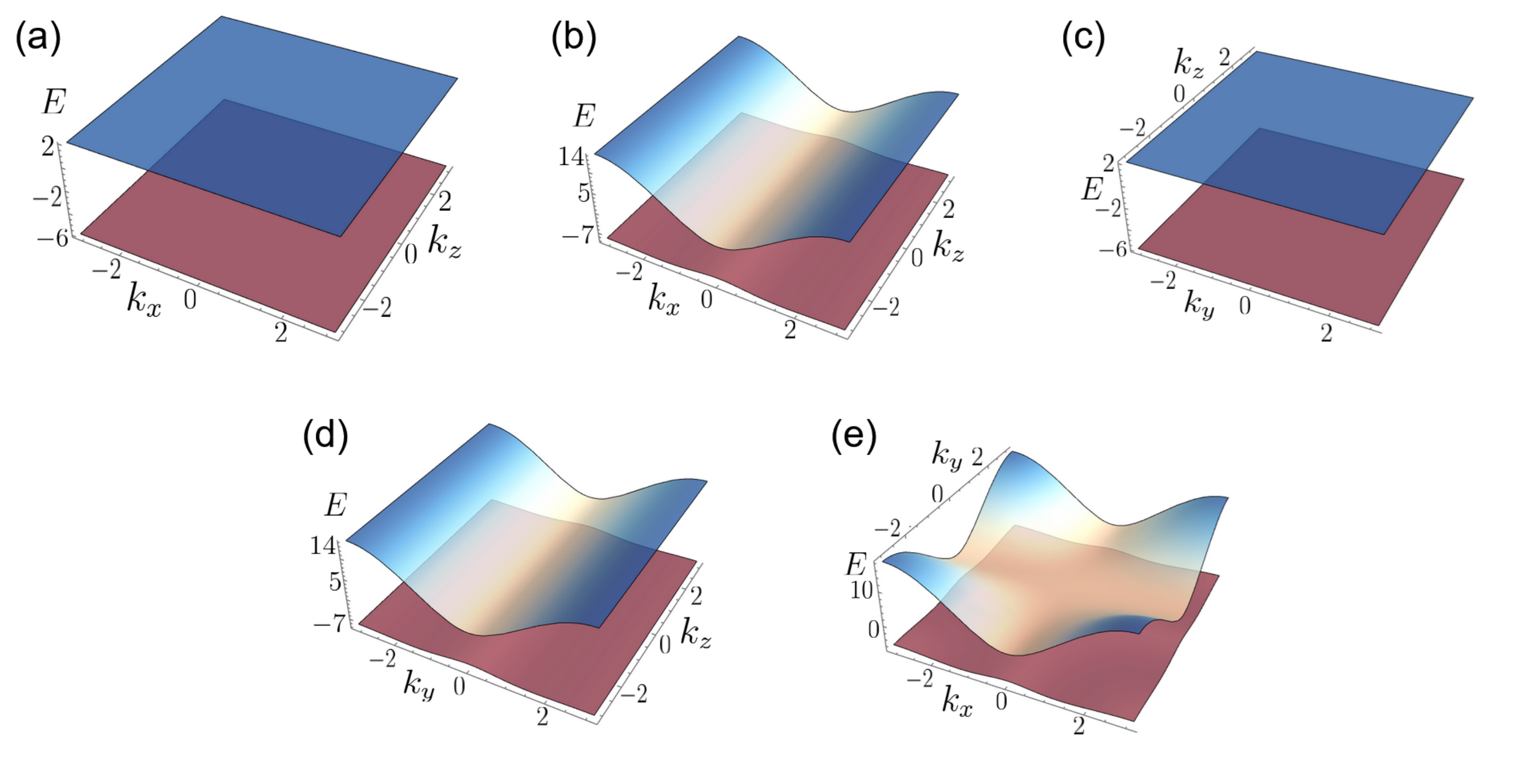}
	\caption{ Energy spectrum of Eq.~(\ref{eq:E_3Dky}) for a) $k_y=0$, b) $k_y=2$, c) $k_x=0$, d) $k_x=2$, and e) any $k_z$ planes for $h=-(t_a-t_b)=2t_2$, $4t'=-(t_a+t_b)$, $h = -4$, $t' = 0.5$ and $t_2 = -2$. As discussed in the main text, both the bands are exactly flat for $k_x=0$ and  $k_y =0$ planes. In contrast, for other 2D plane (any $k_z$) one of the bands become dispersive while the other becomes nearly flat.}
	\label{fig:3D_flat_bands}
\end{figure}

Next, we investigate the presence of flat bands in $H_{3D}$. Notice that although $d_x({\bf k})$ and $d_y({\bf k})$ individually depends on $k_z$, the square of their sum $d_x({\bf k})^2+d_y({\bf k})^2$ is independent of $k_z$. The $k_z$ dependence enters in the spectrum only through $t_3$ term in $d_z({\bf k})$. However, for $t_3\ll t_a, t_b, t', t_2$, the bands are nearly flat along $k_z$ and dispersive only along $k_x$ and $k_y$ for a particular configuration of parameters. For simplicity, we set $t_3=0$, and this simplifies energy spectrum in the $k_y=0$ plane as
\begin{align}
E(k_x,0,k_z)=&(t_a + t_b) + (t_a + t_b + 4 t') \cos(k_x) \pm[h^2 + \frac{3}{2}(t_a - t_b)^2+ 2 t_2^2 + 2 h (t_a - t_b)\nonumber\\
&+ \frac{1}{2}(t_a - t_b - 2 t_2)
(t_a - t_b + 2 t_2 )\mbox{cos}(2k_x) + 2 (t_a - t_b)(t_a - t_b+h)\mbox{cos}(k_x)]^{1/2}
\label{eq:E_3Dky}
\end{align}
Remarkably, for the same flatness conditions in the 2D case, we obtain exact flat bands (see Fig.~\ref{fig:3D_flat_bands}a) with energies 
\begin{align}
E(k_x,k_z)=(t_a+t_b)\pm h.
\end{align}
For $k_y\ne 0$  planes, one of the bands becomes dispersive while the other one remains nearly flat as shown in Fig.~(\ref{fig:3D_flat_bands}b). This also applies to $k_x=0$ and $k_x\ne 0$ planes. Fig. ~(\ref{fig:3D_flat_bands}c-d) demonstrates these features. In contrast to $k_x=0$ and $k_y=0$ planes, we obtain one band nearly flat in any $k_z$ planes with the same parameter discussed in the 2D case. The corresponding plots are shown in Fig.~(\ref{fig:3D_flat_bands}e). We note that for finite but small $t_3$, the flatness of the bands retains.

\begin{figure}
\centering
	\includegraphics[width=0.65\linewidth]{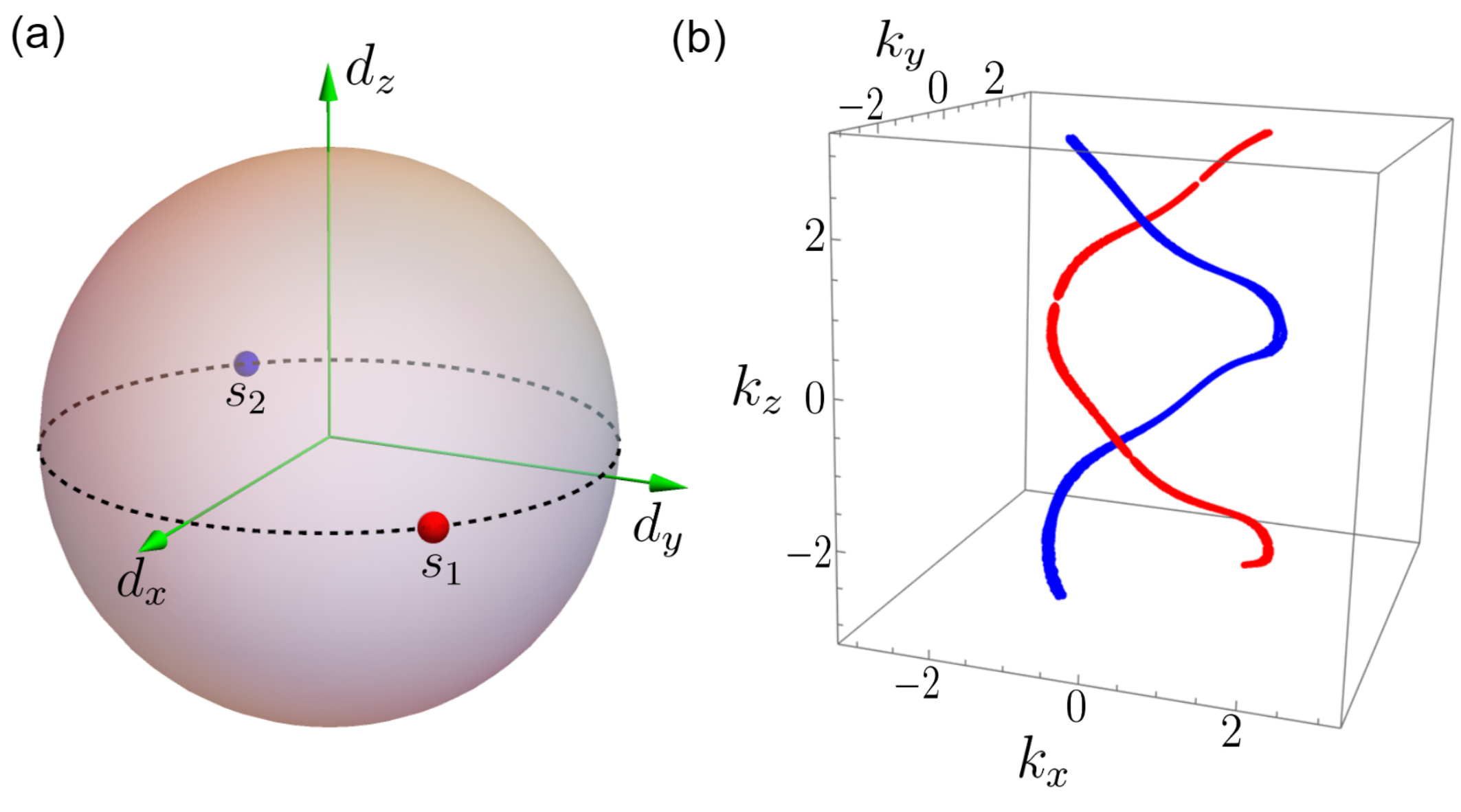}
	\caption{(a) Plot of Bloch sphere with unit radius. Red and blue points represent $s_1$ and $s_2$ as discussed in the main text. These two points are equated with the ${\bf d({\bf k})}$ to find the solution ${\vec k}$. The solutions are then plotted in (b). The parameters are same as in Fig.~\ref{fig:2D_Model_Plot}.}
	\label{fig:hopf_invariant}
\end{figure}

\section{Topological Invariant}
Having flat bands in a genuinely 3D cubic lattice model, we now characterise these flat bands either by topological Chern invariant or by Hopf invariant. We first compute Chern number $C$ for different flat bands discussed in the preceding sections. While flat bands in the $y-z$ and $x-z$ planes carry zero Chern number irrespective of the values of the parameters of the Hamiltonian, the flat band in the $x-y$ plane has finite Chern number similar to the 2D case discussed before. Thus the 3D model possesses a finite Chern number in the $x-y$ plane.

We now find if such a 3D model can host 3D Hopf invariant. To measure it, we introduce a Bloch sphere for the Hamiltonian $H_{\textrm 3D}={\bf d({\bf k})}\cdot {\bf\sigma}$ in Eq.~(\ref{eq:3DMomentum_Ham}). The first term $d_0(\textbf{k})$ configures only the band spectrum but has nothing to do in classifying the topology of the system. Here for every {\bf k}-point in the 3D Hamiltonian, we can identify a specific Bloch vector on the unit sphere defined as, $\hat{\textbf{d}}_{\textbf{k}} = \textbf{d}(\textbf{k})/|\textbf{d}(\textbf{k})|$. Any unit vector in the spherical-polar coordinate can generically be written in terms of azimuthal ($\phi$) and polar ($\theta$) angle by
\begin{align}
\hat{\textbf{d}}_{\textbf{k}} =  (\sin(\theta_{\textbf{k}})\cos(\phi_{\textbf{k}}), \sin(\theta_{\textbf{k}})\sin(\phi_{\textbf{k}}), \cos(\theta_{\textbf{k}})).
\end{align}

The 3D Brillouin zone (BZ) can be identified as a 3-torus ($\textbf{T}^3$) and Bloch sphere of unit radius ($\hat{\textbf{d}}_{\textbf{k}}$) as 2-sphere ($\textbf{S}^2$). Following Ref. {\cite{PhysRevA.99.043619}, we define a map $f:\textbf{T}^3 \rightarrow \textbf{S}^2$. Note that there is a dimensional reduction between the initial and final manifold space. This implies there exists a one-dimensional sub-manifold in BZ for the preimage $f^{-1}(s)$, where $s \in \textbf{S}^2$ is a zero-dimensional point on the Bloch Sphere.  For any two point $s_1 , s_2 \in \textbf{S}^2$, the linking number between the preimage $f^{-1}(s_1)$ and $f^{-1}(s_2)$ is related with the Hopf Index of the system.

To visualize this, we consider two antipodal points $s_1 = (1/\sqrt{2},1/\sqrt{2},0)$ and $s_2 =(-1/\sqrt{2}, -1/\sqrt{2},0)$ from Bloch sphere in its equatorial plane($\theta_{\textbf{k}} = \pi/2$) as shown in Fig.~(\ref{fig:hopf_invariant}). The simultaneous solution of the three equations 
\begin{align}
\{d_{x}({\bf k}),d_{y}({\bf k}),d_{z}({\bf k})\} = \{\frac{1}{\sqrt{2}},\frac{1}{\sqrt{2}},0\}
\end{align}
for the point $s_1$ gives a one-dimensional submanifold in the BZ. The red line in Fig.~(\ref{fig:hopf_invariant}) shows the solution and its $\epsilon$-neighbourhood for $f^{-1}(s_1)$, where $\epsilon \ll 0.1$ and $|f^{-1}(s_1)_{plot}| - |f^{-1}(s_1)_{exact}| \le \epsilon$. Similarly, if we solve for point $s_2$ and plot the blue line, it is observed that one line winds around other once. This implies the magnitude of Hopf index is 1.

\begin{figure} 
\centering
	\includegraphics[width=0.87\linewidth]{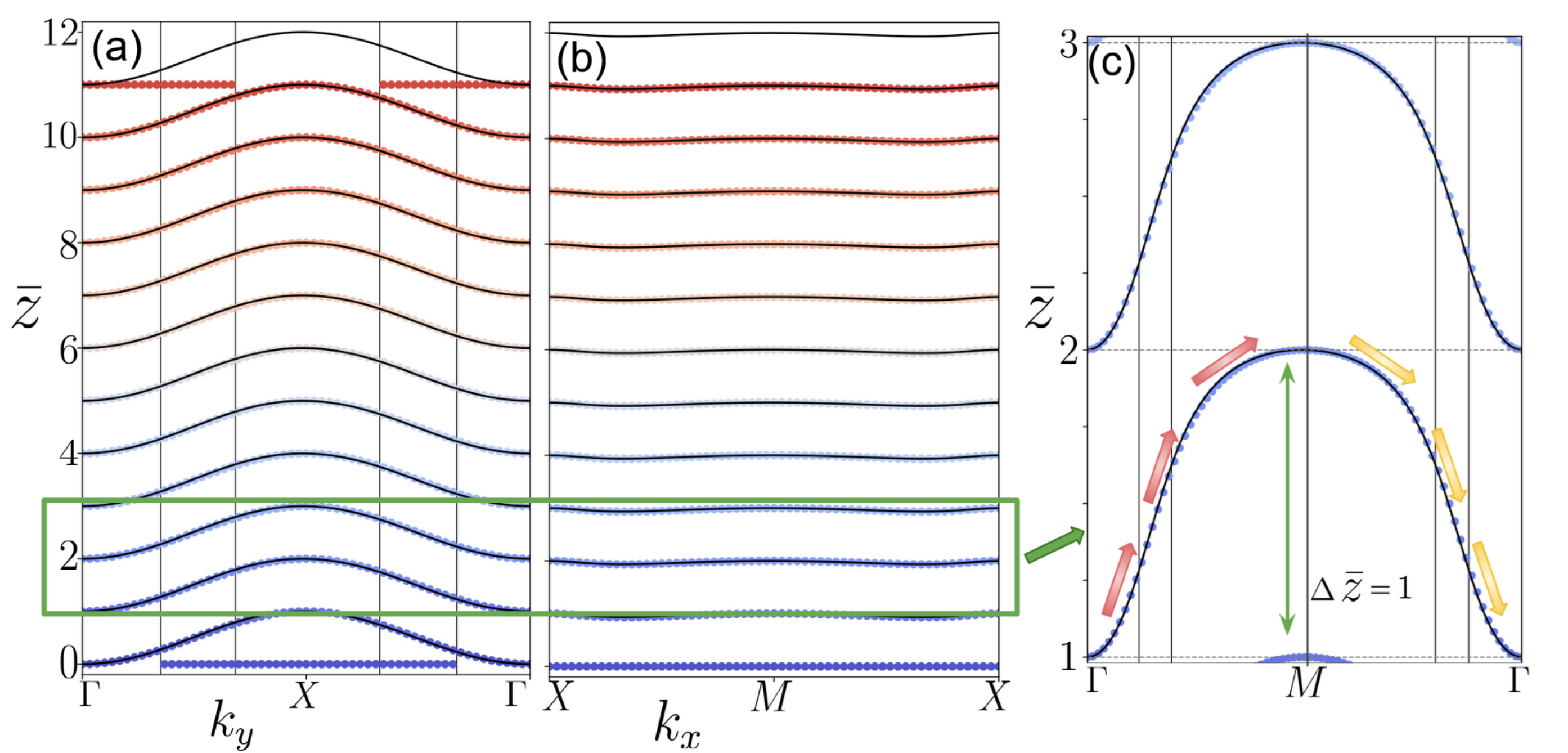}
    \includegraphics[width=0.87\linewidth]{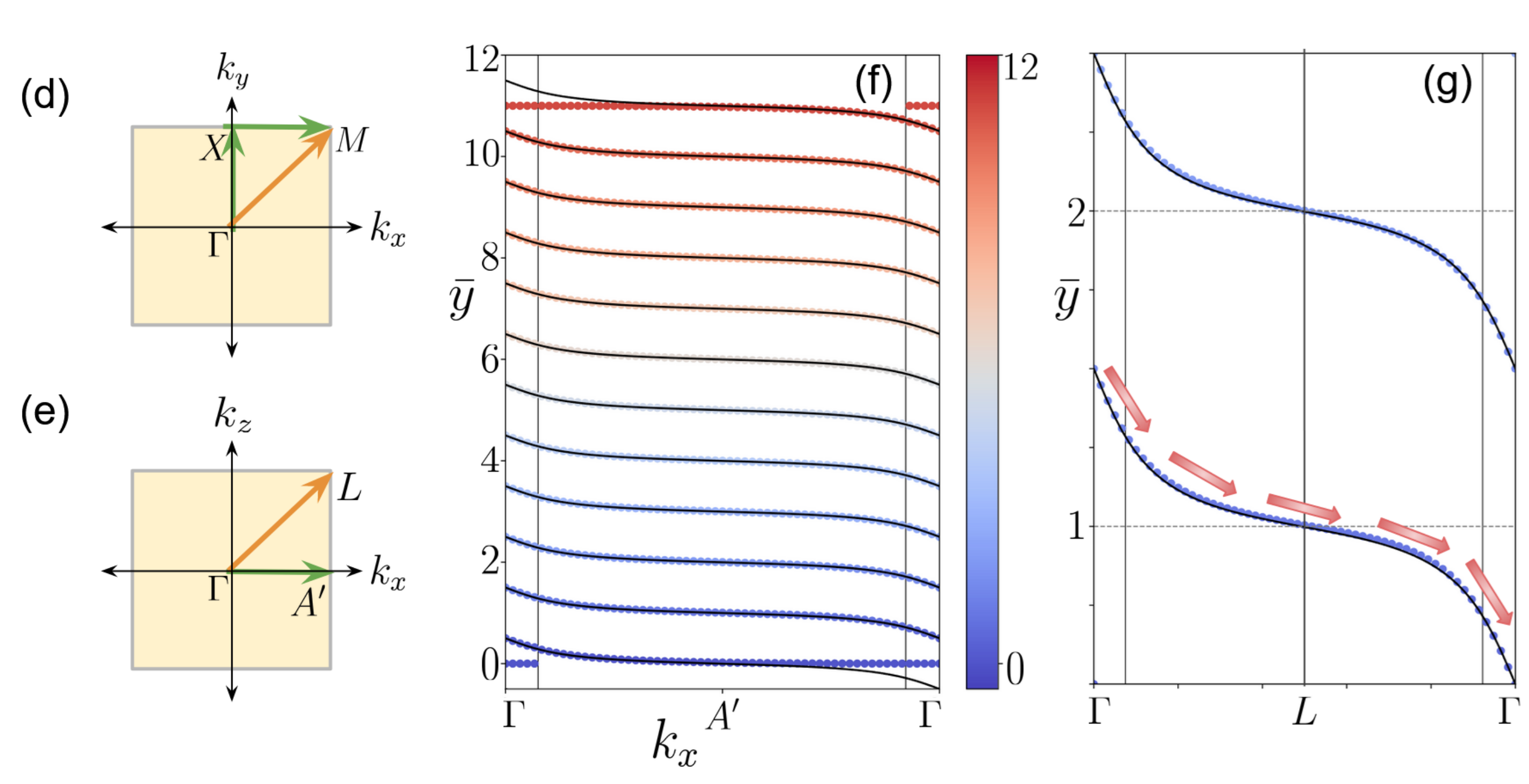}
	\caption{Plots of Hybrid Wannier Function Centers (HWFC) \cite{PhysRevLett.102.107603} in different planes of Hopf-Chern Insulator. The black continuous line illustrates the bulk HWFC, while layered HWFCs are represented by colored dots. The color map indicates the position expectation value for a 12-layer slab. (a) HWFC computed along z in the rBZ ($k_x-k_y$) along $\Gamma (0,0)$ and $X (0, \pi)$ points reveals that the charge center translates precisely by one lattice period at the halfway mark before returning to its original position. (b) Similarly, the plot along $X (0, \pi)$ to $M (\pi, \pi)$ shows the translation pattern of the HWFC, where it is localized in the vicinity of it's layer. (c) A snapshot of HWFC for the z-th layer 1 and 2, to clearly observe that $\Delta z = 1$ in between the $C_{4z}$ rotation symmetry invariant points $\Gamma$ and $M$. (d) The paths in the $k_x-k_y$ Brillouin zone along which a, b and c are plotted. (e) The paths in the $k_x-k_z$ Brillouin zone along which f and g are plotted.   (f) The HWFC computed along y in the rBZ ($k_x-k_z$) along $\Gamma (0,0)$ and $X (\pi, 0)$ points demonstrate that the charge center undergoes a translation of exactly one lattice period in its full path. (g) A snapshot of HWFC for the y-th layer 1 and 2, to clearly observe that HWFC traverses from $[1 0 0]$ plane to $[\bar{1} 0 0]$ as a consequence of Wannier obstruction.}
	\label{fig:RTP}
\end{figure}
In addition to the Hopf Index, we can characterise the present model by another topological invariant, namely Returning Thousless pump (RTP) invariant. This is associated with the nature of the hybrid Wannier function centers (HWFCs) of the corresponding lattice model. The HWFCs in real space are also important as it show a novel localized property not typically observed in conventional topological insulators. While for the standard Chern insulators, the Wannier functions are extended in nature \cite{PhysRevLett.98.046402}, they are exponentially localized across multiple unit cells \cite{PhysRevB.103.045107, PhysRevLett.126.216404, PhysRevB.106.075124} in the direction perpendicular to the Chern plane for the present model. To elucidate this feature, we take two different choices of Hybrid Wannier Function (HWF). For the first case, the $\hat{z}$ direction (perpendicular to the Chern plane) of the lattice model is Wannier-like, and the $k_x$-$k_y$ plane is Bloch-like, forming the HWF defined as: $\ket{W_{nz}(k_x, k_y)} = \frac{1}{2\pi} \int dk_z\, e^{i\mathbf{k} \cdot (\mathbf{r} - z\hat{z})}$. In the second case, the HWF is Wannier-like in the $\hat{y}$ direction of Chern plane. We plot the hybrid Wannier functions centers (HWFC) $\bar{r}_i$ ($i \in \{x,y,z\}$) for these two different choices following the theory illustrated in Appendix \ref{RTP}. These centers are related to the Berry phase as $\bar{r}_i(k_j, k_k) = \int_0^{2\pi}\, dk_i A_i(\mathbf{k})/2\pi$, where $A_i$ is the Berry connection and thereby linked with the Thouless charge pumping ($\mathscr{P}$) according to the geometric theory of polarization \cite{PhysRevLett.102.107603}. For the Hopf-Chern case, the HWFCs exhibit both localized and extended nature simultaneously in different crystal planes as illustrated in Fig.~(\ref{fig:RTP}).}

In Figs.~(\ref{fig:RTP}a) and (\ref{fig:RTP}b), the HWFCs are illustrated for a 12-layered slab stacked along the $\hat{z}$ direction with inter-layer hopping. It is evident that the position center shifts from its original layer to the unit layer above along the $C_{4z}$ rotation symmetric point (Fig. \ref{fig:RTP}a, b, c) from $\Gamma$ to $M$ in the BZ. The path of these plots in the BZ is shown in Fig. (\ref{fig:RTP}d). The spectral flow plot in Fig. (\ref{fig:RTP}c) clearly demonstrates that the HWFC spreaded from 1 to 2 in the 1st half of BZ (or more generally from $z$ to $z+1$, as indicated by the red arrow) returns in the second half to its original position (as depicted by the yellow arrow), suggesting a periodic return to the initial layer. This implies there is no topological obstruction as observed in the stable topological insulators though the Wannier function is spreaded over multiple unit cells.

This kind of scenario is in contrast to the conventional Wannier obstruction observed in the Chern plane, as clearly shown in Figs.~(\ref{fig:RTP}f) and (\ref{fig:RTP}g). For this case, the HWF is Wannier-like in the $\hat y$ direction and intersects the Chern plane. Consequently, the position center of the HWF, $\bar{y}_n(k_x, k_z)$, spans a full unit layer and does not return to its original position, as depicted by the spectral flow in Fig.~ (\ref{fig:RTP}g). This indicates that the Wannier function is not localized in the vicinity of its layer and is extended over the Brillouin Zone, highlighting the effect of topological obstruction. The spectral flow of the HWFC is related to the traditional Thouless charge pumping \cite{PhysRevB.27.6083} in terms of topology, implying that the charge is pumped from the $(010)$ to the $(0\Bar{1}0)$ surface layer in the BZ. This is in contrast to the first case where the charge is pumped from the $z$ to $z+1$ th layer in the first half, and then it returns to the original layer in the second half. This phenomenon is characterised as Returning Thouless Pump as mentioned in literatures \cite{PhysRevB.103.045107,  PhysRevB.106.075124}. Accordingly, the quantized difference in HWFC in the $\hat{z}$ direction between two $C_{4z}$ invariant points defines the RTP invariant. It is important to emphasize that Hopf invariant and RTP invariant are distinct concepts. Even these two topological features may not necessarily come together in any specific lattice models. While the RTP invariant is a delicate topological invariant protected by crystalline rotational symmetry, the Hopf invariant can also survive even when rotational symmetry in the crystal is relaxed.  In the present $C_{4z}$ rotational symmetric model, the RTP invariant turns out to be $\Delta \mathscr{P}_{\Gamma X} = 1$ as evident from Fig.~(\ref{fig:RTP}a). However, along $X\rightarrow M$, $\Delta \mathscr{P}_{XM} = 0$ as the centers are confined within each layer (cf. Fig.~\ref{fig:RTP}b). Consequently, we obtain $\Delta \mathscr{P}_{\Gamma M} = 1$ (Fig.~\ref{fig:RTP}c).

\section{Phase diagram} 

\begin{figure}
\centering
	\includegraphics[width=0.59\linewidth]{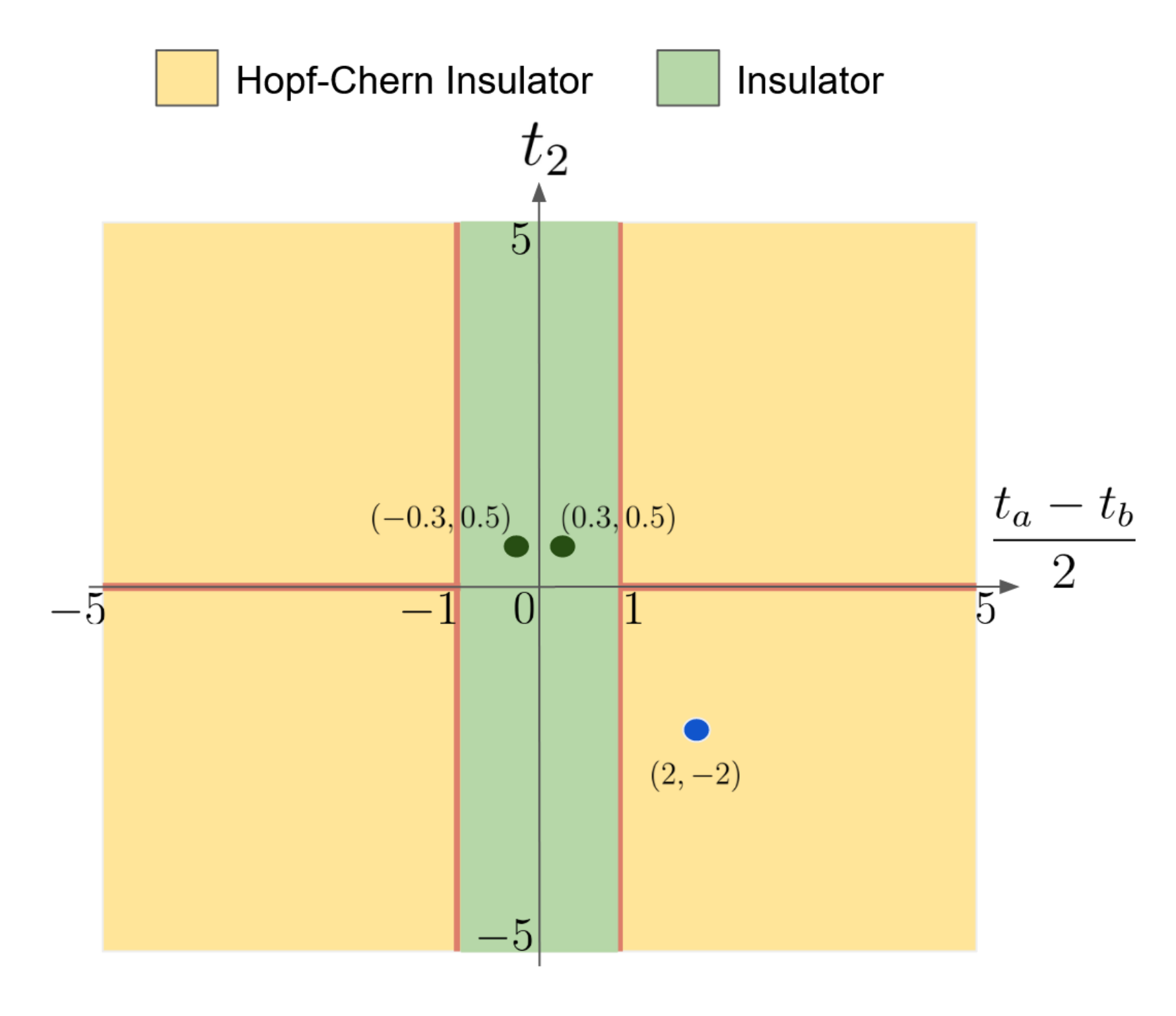}
	\caption{Phase diagram of 3D Hamiltonian in the plane $t_2$-$(t_a-t_b)/2$  with $h=-4$, $t'=0.5$ and $(t_a+t_b)=-2$. Note that the Hopf-Chern phases are symmetric with respect to $(t_a-t_b)=0$. The bulk spectrum closes at red colored line. The blue dot represents parameters for having flat bands in the Hopf-Chern phase whereas the dark green dots represent parameters for flat bands in the trivial insulating regime.}
	\label{fig:phases}
\end{figure} 

Equipped with the computation of Hopf invariant, we now chart out the phase diagram of the Hamiltonian in Eq.~(\ref{eq:3DMomentum_Ham}). Fig.~(\ref{fig:phases}) illustrates several distinct phases in the parameter space $t_2$ and $(t_a-t_b)$ with $h=-4$, $t'=0.5$ and $(t_a+t_b)=-2$. Evidently, the Hopf-Chern phases appear symmetrically in the $t_2$-$(t_a-t_b)/2$ plane. For $ 2 |(t_a-t_b)|> |h|$, we find insulating phase with finite 3D Hopf and 2D Chern invariants. Otherwise, we obtain trivial insulating phase.  At the transition point i.e,  $(t_a-t_b)=\pm h/2$ (vertical red colored line in Fig.~(\ref{fig:phases})), the band gap closes at a single $k$ point in the BZ. In the non-trivial regime, the flat bands appear only at a special configuration of parameters (blue region) as indicated in Fig.~(\ref{fig:phases}). It is worth pointing out that the trivial insulating regime may also host flat bands in different planes. For a more weaker orbital-flipping NNN hopping $t_2=0.5$, the nearly flat band with a flatness $\sim 1/21$ of band gap in plane $k_x\sim 1$ or $k_y\sim1$ is obtained for the parameters $h=-4$, $t'=0.5$, $(t_a+t_b)=-2$ and $(t_a-t_b)=\pm 0.6$. This is denoted by dark green dots in the phase diagram.
\section{Higher Chern model and Flatness}
\begin{figure}
\centering
\includegraphics[width=0.65\linewidth]{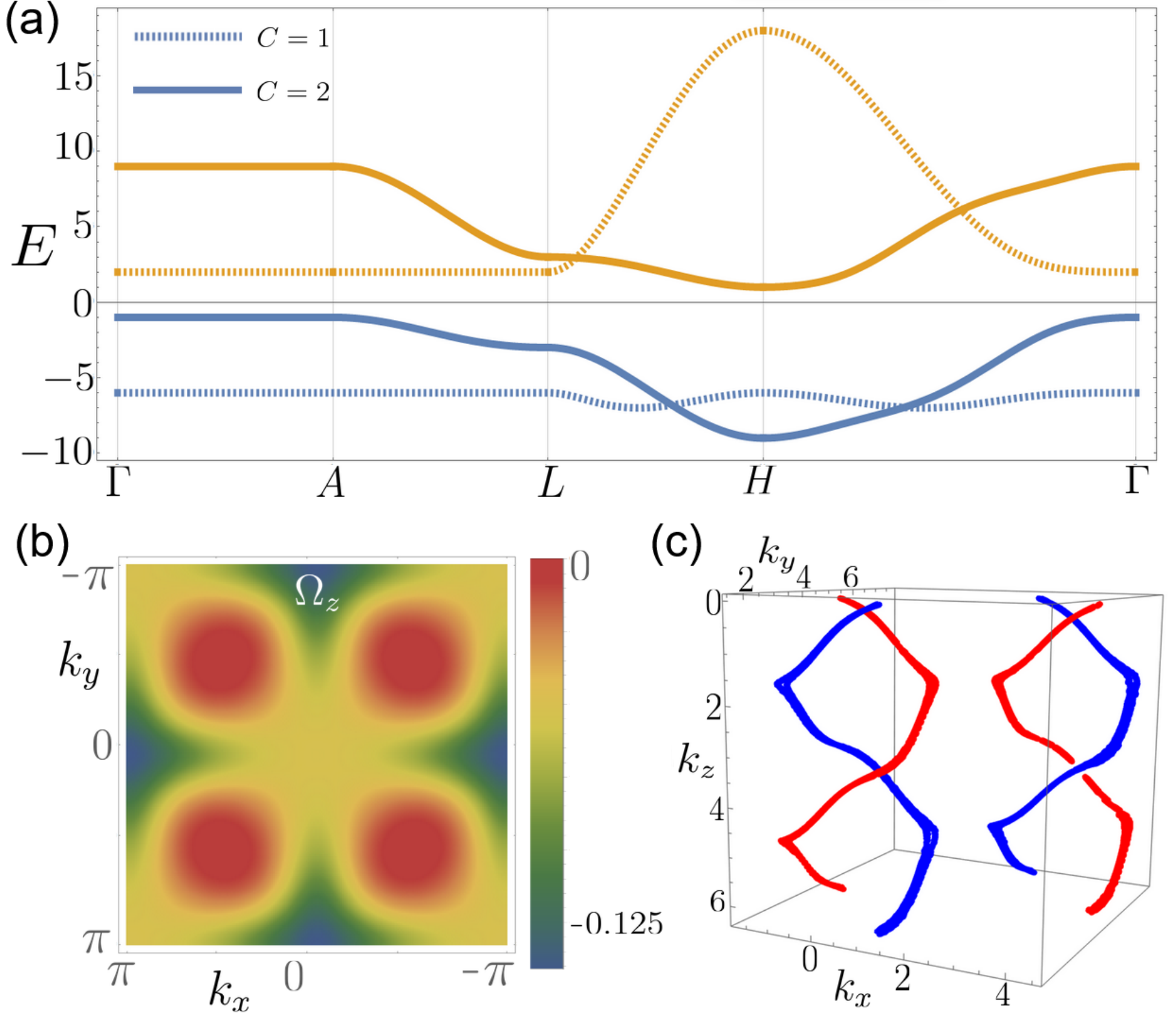}
	\caption{(a) Bulk band spectrum along the high symmetric path in the cubic BZ for both $C=1$ and $C=2$ phases. The parameters for the $C=2$ phase are taken to be $h=1$, $t_a=t_b=1$, $t'=1$, $t_2=-2$ and for $C=1$ phase we use parameters as in Fig.~\ref{fig:3D_flat_bands}. (b) Berry curvature plot for C = 2 phases. (c) The winding of preimages in the Hopf-Chern model with Chern number $C=2$.  }
	\label{fig:hopf_Chern2}
\end{figure}

We next construct a 3D Hopf model which exhibits higher Chern phases with only NN and NNN as before. We take $t'_{AA}=-t'_{BB}=t'$ in Eq.~(\ref{eq:hoppings}). This leads to $d_0(k)=(t_a+t_b)(\cos k_x+\cos k_y)$ and $d_z (k)=(h+(t_a-t_b)(\cos k_x+\cos k_y))+4t'\cos k_x \cos k_y$. $d_x$ and $d_y$ remain same as before. For the region $-(h/2 +2) < (t_a - t_b) < (h/2 +2)$, the model exhibits Chern phases with $C=2$ for $-4<h<4$ as illustrated in Appendix \ref{C2}. With this, the 3D model can be constructed following the same procedure discussed in the preceding sections. To find Hopf invariant, the preimages of two antipodal points on the Bloch sphere are shown in Fig.~(\ref{fig:hopf_Chern2}c) for $h=1$, $t'=1$, $t_2=-2$ and $t_a=t_b=1$. Evidently, there is a winding between the two preimages, confirming a 3D model of both Hopf and higher Chern invariants with NN and NNN hopping. Moreover, the presence of two loops indeed indicates the two gaped Dirac points in the spectrum responsible for $C=2$ phase. It is worth pointing out that finding Hopf invariant in a higher Chern model is subtle and somewhat it depends on the location of Dirac point within the BZ. For example, the winding of preimages turns out to be easily identifiable in a Chern model with gaped Dirac points at the center of the BZ. In contrast, Chern models with gapped Dirac nodes at arbitrary momenta often give solutions for preimages for which number of winding is not easy to count. Hence, a better diagnostic is required which is beyond the scope of the current work.

\par We further comment if such a 3D Hopf-Chern model can host flat bands. Evidently, $d_0$ is now a constant parameter as $t_a$ and $t_b$ are fixed to allow only $C=2$ phase in the 2D model in Eq.~(\ref{eq:2D_Bulk_Ham}). Consequently, the 3D model does not support flat bands along the high-symmetry path in the BZ as evident in Fig.~(\ref{fig:hopf_Chern2}). For comparison, we have also added band spectrum for $C=1$. To get flat bands in such a 3D model, we must incorporate higher-order hopping which seems to be a natural choice (we do not present here for simplicity). Thus, finding flat bands in a higher Hopf-Chern model with NN and NNN neighbor is not guaranteed.

\section{Conclusion}
In conclusion, we study flat bands with non-trivial Hopf and Hopf-Chern number in 3D lattice models. We find that a pure Hopf model can host perfect flat band without incorporating any long-range hopping beyond its usual range of hopping. We further show a lattice model exhibiting both Hopf and Chern invariant can be constructed from a 2D Chern insulator. Depending upon the parameters of the model Hamiltonian and crystal planes, completely or nearly flat bands are obtained. We further show that a 3D model with higher Chern number is realizable with only NN and NNN hopping in the same setting, but the presence of flat bands in all crystal planes is not guaranteed. We point out that it is indeed an open problem to explore if any Chern model with arbitrary Chern number can be converted to a Hopf model with flat bands. This requires further investigation. In addition, since real materials involve many bands in their low energy spectra, it is desirable to find flat bands in multiband Hopf models.

\par The lattice model proposed here is expected to be realizable in three-dimensional optical lattices of dipolar interacting spins. It has already been proposed that the standard Hopf insulators can be realized in polar molecules loaded in 3D optical lattices where dipolar interaction can be tuned to implement various hopping\cite{PhysRevLett.127.015301, PhysRevA.103.063322}. In addition, Hopf model has been shown to be realizable in a circuit\cite{PhysRevLett.130.057201}.  Since the proposed set up in the current work requires only short range hopping than the standard Hopf model\cite{PhysRevLett.101.186805}, we believe that the flat bands and related Hopf physics can easily be achieved in 3D optical lattices with dipolar interaction and in an electric circuit. 

\section{Acknowledgement} We thank Shamik Banerjee for useful discussion. ID gratefully acknowledges Alexei Andreanov, Bohm-Jung Yang and the participants of the Intercontinental Binodal Workshop on {\it Flat Bands and High-Order Van Hove Singularities} at PCS, IBS, Daejeon, for their valuable discussions during the manuscript revision process. KS acknowledges funding from the Science and Engineering Research Board (SERB) under SERB-MATRICS Grant No. MTR/2023/000743.

\section{Appendix}
\subsection{\label{Hopf Flat} Hopf Tight-binding Model with Exact Flat Band}
The tight-binding Hamiltonian of a two-orbital cubic lattice with nearest neighbour (NN), next-nearest neighbour (NNN) and next-to-next-nearest neighbour (NNNN) hopping can be expressed as 
\begin{align}
H_{\textrm {hopf}} = \sum_{\Vec{\mathbf{r}}, \mu, \nu} \ket{\Vec{\mathbf{r}} + \Vec{e_{\mu}}}\:\:\Gamma^{\mu\nu}\:\:\bra{\Vec{\mathbf{r}} + \Vec{e_{\nu}}}
\label{eq:Hopf_TBA}
\end{align}
Here $\mu$,$\nu$ are characterized by their position in the array of hopping direction $(0,x,y,z,xy,yz,zx,2x,2y,2z)$. $\Vec{e}_{\mu/\nu}$ is the position vector of the hopping. As for example, $\Vec{e}_{0} = (0,0,0)$, $\Vec{e}_{4} = (1,1,0)$ and $\Vec{e}_{9} = (0,0,2)$. The Hopf model without any Chern phases requires the following $\Gamma$'s to be nonzero: 
\begin{alignat}{3}
\Gamma^{00}&=  -(1+ h^2) \sigma_z; & \Gamma^{10}&= h (-i\sigma_y - \sigma_z);& \Gamma^{20}&= h  (-i\sigma_x - \sigma_z);\nonumber\\
\Gamma^{30}&= -h\sigma_z; & \Gamma^{40}&= \frac{1}{2}(- i\sigma_x -i \sigma_y - \sigma_z);\quad \quad&\Gamma^{50}&= \frac{1}{2}(- i\sigma_x + \sigma_y - \sigma_z);\nonumber \\
\Gamma^{60}&= \frac{1}{2}(- \sigma_x -i \sigma_y - \sigma_z);\qquad & \Gamma^{70}&= \frac{1}{2}(-i \sigma_y - \sigma_z); &\Gamma^{80}&= \frac{1}{2}(- i\sigma_x  - \sigma_z); \\
\Gamma^{12}&= \frac{1}{2}( i\sigma_x -i \sigma_y - \sigma_z); & \Gamma^{23}&= \frac{1}{2}(- i\sigma_x - \sigma_y - \sigma_z); &\Gamma^{13}&= \frac{1}{2}( \sigma_x - i \sigma_y - \sigma_z) \nonumber
\label{eq:Gamma}
\end{alignat}
Clearly, the pure Hopf requires NN, NNN and NNNN hoppings. The Hopf invariant for $-3<h<3$ is obtained to be $\mathbb Z$. Interestingly, without introducing any additional range of hopping, we obtain one band with $E=0$. This requires the following terms to add in Eq.~(\ref{eq:Hopf_TBA}).
\begin{alignat}{3}
\Gamma^{00}&= (3 + h^2)\mathbb{1};\quad\quad \quad& \Gamma^{10}&= h \mathbb{1}\qquad\qquad& \Gamma^{20}&= h\mathbb{1};\nonumber\\
\Gamma^{30}&= h\mathbb{1} ; & \Gamma^{40}&= \frac{1}{2}\mathbb{1};&\Gamma^{50}&= \frac{1}{2}\mathbb{1};\nonumber \\
\Gamma^{60}&= \frac{1}{2}\mathbb{1}; & \Gamma^{70}&= 0; &\Gamma^{80}&= 0; \\
\Gamma^{12}&= \frac{1}{2}\mathbb{1}; & \Gamma^{23}&= \frac{1}{2}\mathbb{1}; &\Gamma^{13}&= \frac{1}{2}\mathbb{1} \nonumber
\label{Gamma1}
\end{alignat}

\subsection{\label{Hopf-CLS} Compact Localized State of Pure Hopf Model}

The compact localized state (CLS) represents an optimum case of Wannier function, characterised by nonzero amplitude confined to a finite region and zero amplitude elsewhere. This property encapsulates the localized nature of a flat band in real space, as a consequence of destructive interference due to the specific geometry of lattice and hopping patterns of electrons. Given the degenerate Bloch eigenfunctions corresponding to flat bands, it is possible to combine them to form new eigenfunctions that exhibit CLS within a bounded region. Following the ref \cite{PhysRevB.99.045107}, the n-th Bloch eigenfunction of bipartite Hamiltonian $H(\vec{k})$ corresponding to the flat band energy $E_n$ and eigenvector ${\bf v}_n(\vec{k})$ can be defined as,
\begin{equation}
    \ket{\psi_n({\vec{k}})} = \frac{1}{\sqrt{N}} \sum_{\vec{R}} \sum_{q = 1}^2 \, v_{n,q}(\vec{k})\, e^{i \vec{k}.\vec{R}} \, \ket{\vec{R},q} 
\end{equation}
Then the new eigenfunction in the real space as a linear combination of Bloch eigenfunctions can be expressed as,
\begin{equation}
\begin{split}
    \ket{\text{CLS}_R} =& \, C \sum_{\vec{k} \in BZ} \, \alpha (\vec{k}) \, e^{-i \vec{k}.\vec{R}} \ket{\psi_n({\vec{k}})}\\
    =& \, \frac{C}{\sqrt{N}} \sum_{\vec{R'}} \sum_{q = 1}^2 \sum_{\vec{k} \in BZ} \, \alpha (\vec{k}) \, v_{n,q}(\vec{k}) \, e^{i \vec{k}.(\vec{R'} - \vec{R})}\, \ket{\vec{R'},q}\\
    =& \sum_{\vec{R'}} \sum_{q = 1}^2 {\bf M}_{\vec{R}, \vec{R'}, q}\, \ket{\vec{R'},q}.
\end{split}
\end{equation}
Here, $C$ is the normalization constant and $\alpha (\vec{k})$ tunes the amplitude of the linear combination of different Bloch eigenfunctions. The condition for $\ket{\text{CLS}_R}$ to be valid is that, ${\bf M}_{\vec{R}, \vec{R'}, q}$ (the coefficient of  $|\vec{R'},q\rangle$) must be non-zero only in a small bounded region and zero elsewhere, which can be controlled by $\alpha (\vec{k})$ over the BZ. It can be directly observed that ${\bf M}_{\vec{R}, \vec{R'}, q}$ is just an inverse Fourier transform of $\alpha (\vec{k})\, v_{n,q}(\vec{k})$. This implies that if $\alpha (\vec{k})\, v_{n,q}(\vec{k})$ is only a finite polynomial of Bloch phases, the CLS is guaranteed. Even for Hopf flat Hamiltonian, this can be achieved by the unnormalized eigenstate corresponding to the flat band energy.

For Hopf Hamiltonian, the eigenvector corresponding to the flat band can be evaluated as,
\begin{equation}
v(\vec{k}) = N(\vec{k}) \begin{pmatrix}
h + \cos k_x + \cos k_y + e^{-i k_z} \\
i \sin k_x - \sin k_y
\end{pmatrix}
\end{equation}
with the normalization constant, $N(\vec{k}) = [3 + h^2 + 2 h (\cos k_x + \cos k_y + \cos k_z) + 2 (\cos k_x \cos k_y + \cos k_x \cos k_z + \cos k_y \cos k_z)]^{-1/2}$. Choosing $\alpha (\vec{k}) = N(\vec{k})^{-1}$,
\begin{equation}
\alpha (\vec{k})\, v(\vec{k}) \propto \begin{pmatrix}
h + \frac{1}{2} (e^{i k_x} + e^{-i k_x} + e^{i k_y} + e^{-i k_y}) + e^{-i k_z} \\
\frac{1}{2}(e^{i k_x} - e^{-i k_x}) + \frac{i}{2}(e^{i k_y} - e^{-i k_y}).
\end{pmatrix}
\end{equation}
This is clearly a finite sum of Bloch phases, which gives,
\begin{equation}
    {\bf M}_{\vec{0}, \vec{R}} \propto \begin{pmatrix}
2h \delta_{0,0} + \delta_{0,-x} + \delta_{0,x} + \delta_{0,-y} + \delta_{0,y} + 2\delta_{0,z} \\
\delta_{0,-x} - \delta_{0,x} + i\delta_{0,-y} - i\delta_{0,y}.
\end{pmatrix}
\end{equation}
Then the compact localized state around $\bf{0}$ can be expressed as,
\begin{equation}
\begin{split}
    \ket{\text{CLS}_{\bf{0}}} \propto &\, 2h  \ket{0, A} + \ket{+ a_x, A} -  \ket{+ a_x, B} +  \ket{- a_x, A} +  \ket{ - a_x, B}  \\ 
     &\: +  \ket{+ a_y, A}-  i\ket{ + a_y, B} + \ket{ -a_y, A} + i \ket{ -a_y, B} + 2 \ket{ + a_z, A},
    \end{split}
\end{equation}

\begin{figure}[h!]
\centering
	\includegraphics[width=0.43\linewidth]{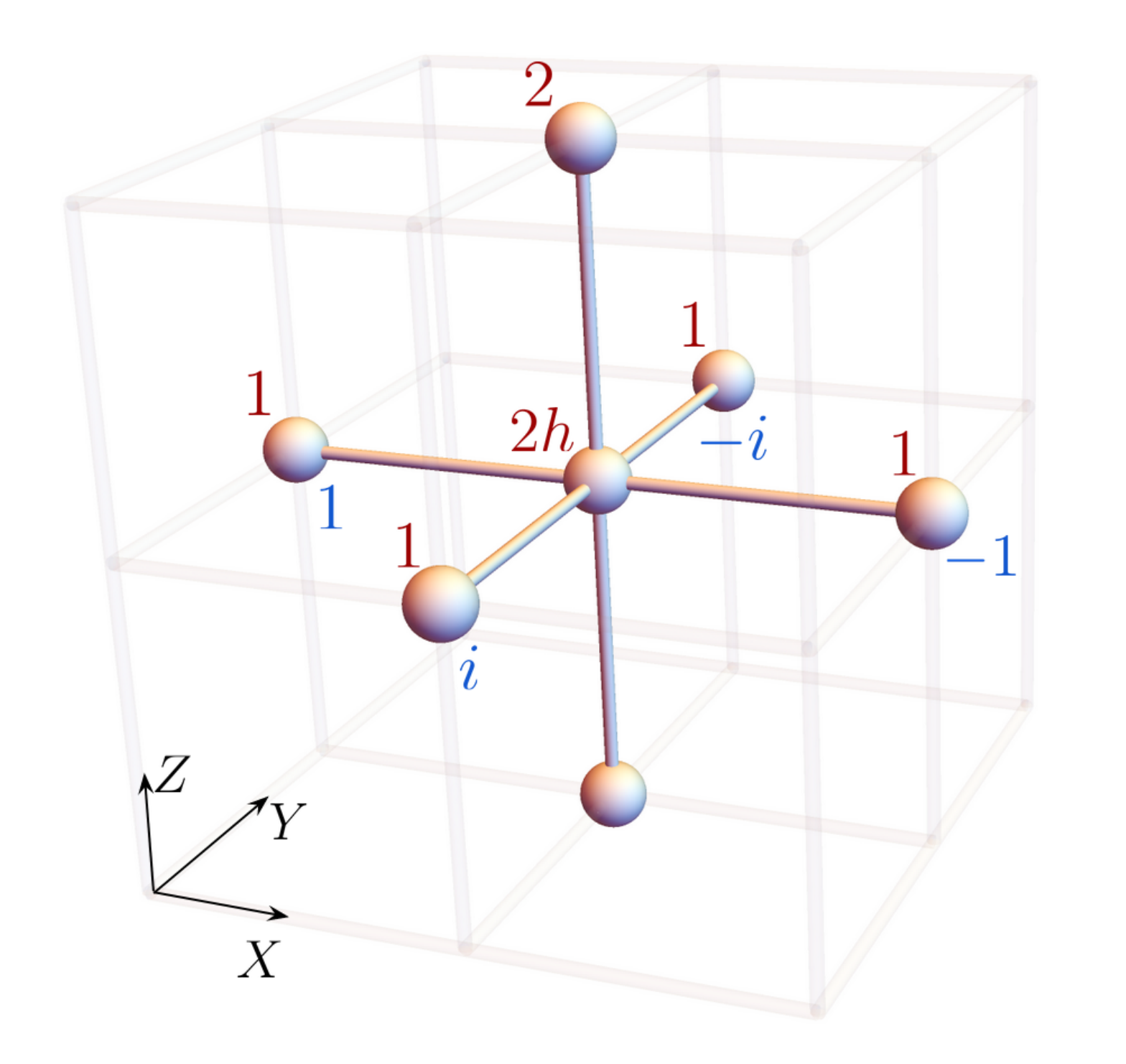}
	\caption{The Compact Localized States (CLS) for pure Hopf Model. The red colour indicates the magnitude for A sublattice and the Blue colour for B sublattice.}
	\label{fig:A0}
\end{figure}
where $a_x$, $a_y$ and $a_z$ are the lattice constants in the three directions. As the normalization constant does not vanish anywhere in the BZ in the Hopf insulator regime, these compact localized states form a complete set.

\subsection{\label{Hopf-Chern Flat} Hopf-Chern Tight-binding Model with Nearly Flat Band}
In a similar spirit, to construct a minimal Hopf-Chern model, we require NN and NNN hopping, where the array of hopping direction is $(0,x,y,z,xy,yz,zx)$. For such a model, the non-zero $\Gamma$ matrices are found to be 

\begin{alignat}{3}
\Gamma^{00}&= h\sigma_z ; & \Gamma^{10}&= \frac{(t_a + t_b)}{2}\mathbb{1} +\frac{(t_a - t_b)}{2}\sigma_z;\quad & \Gamma^{20}&= \frac{(t_a + t_b)}{2}\mathbb{1} +\frac{(t_a - t_b)}{2}\sigma_z;\nonumber\\
\Gamma^{30}&= t_3\sigma_z; & \Gamma^{40}&= t'\mathbb{1};\quad&\Gamma^{50}&= \frac{t_2}{2}(- i\sigma_x + \sigma_y );\nonumber \\
\Gamma^{60}&= \frac{t_2}{2}(- \sigma_x - i\sigma_y );\quad & \Gamma^{12}&= t' \mathbb{1}; &\Gamma^{32}&= \frac{t_2}{2}(i\sigma_x - \sigma_y ); \\
\Gamma^{13}&= \frac{t_2}{2}(\sigma_x - i\sigma_y ) \nonumber
\label{Gamma2}
\end{alignat}
Comparing the $\Gamma$ matrices for Hopf and Hopf-Chern Hamiltonians, it is evident that we need to turn off some hopping terms and tune other hopping parameters to get Hopf-Chern from Hopf lattice model.
\subsection{\label{RTP} Returning Thouless Pump in Hopf and Hopf-Chern Model}
The manifestation of the novel topological features becomes apparent in real-space through the multicellularity of Wannier functions. Traditionally, studies have shown that Wannier functions corresponding to trivial bands exhibit exponential localization. However, recent advancements in the field have revealed a significant departure from this norm: even in the presence of exponential localization, nontrivial bands can emerge, characterized by the expansion of localization across multiple unit cells.

\begin{figure}[h!]
\centering
	\includegraphics[width=0.96\linewidth]{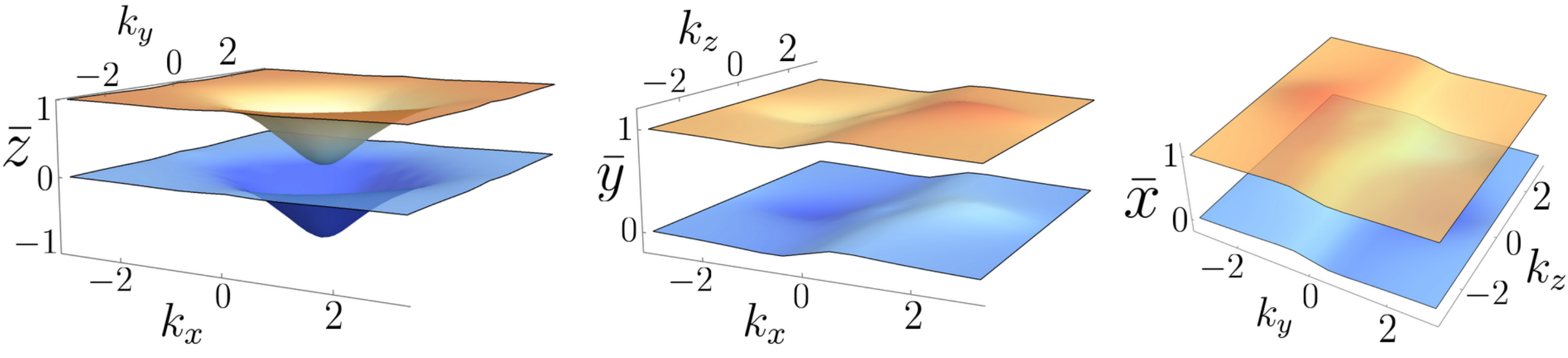}
	\caption{Bulk hybrid Wannier function centres of Hopf Model plotted in the reduced BZ}
	\label{fig:Hopf_WCC}
\end{figure}

The multicellularity phenomenon is observed through the eigenvalues of the projected z-th position operator, $P\hat{z}P$, where $P = \sum_{n,k} \ket{\psi_{nk}}\bra{\psi_{nk}}$ represents the projector onto the bulk-occupied band of the Bloch state $\ket{\psi_{nk}}$. This operator remains invariant under translations in the x-y plane perpendicular to the z-axis. Consequently, the eigenstates can be labeled in the reduced Brillouin zone $k_{\perp} \in (k_x,k_y)$, exhibiting extended nature in the x-y plane while being maximally and exponentially localized in the z-direction. These eigenstates are referred to as hybrid Wannier states. The eigenvalue of this state, $\bar{z}(k_x, k_y) = \bra{W_{n0}(k_x, k_y)}\hat{z}\ket{W_{n0}(k_x, k_y)}$, represents the position center along the z-axis and is illustrated in Figure \ref{fig:Hopf_WCC}.

Translating along the z-axis by a unit lattice constant $P\hat{z}P \rightarrow P(\hat{z}+1)P$ results in an infinite ladder-like structure. Mathematically, the eigenvalue $\bar{z}(k_x, k_y)$ can be expressed as:

\begin{align}
\bar{z}(k_x, k_y)= \bra{W_{n0}(k_{\perp})}\hat{z}\ket{W_{n0}(k_{\perp})} = -\frac{i}{2\pi} \int_0^{2\pi} dk_z \bra{\psi_{nk}}\partial_{k_z}\ket{\psi_{nk}} = \frac{1}{2\pi}\int_0^{2\pi}\, dk_z A_z(\mathbf{k})
\end{align}

This expression demonstrates the relationship between the position operator and the Berry phase, thereby linking it to the polarization ($\mathscr{P}$) according to the geometric theory of polarization.

Since the Hamiltonian exhibits symmetry under rotation, the rotation matrix and $ H(\vec{\mathbf{k}}) $ can be simultaneously diagonalized. Consequently, $ H(\vec{\mathbf{k}}) $ is diagonal along the $ C_{4z} $ symmetry lines $ (\gamma_i) $ with distinct angular momentum sectors. This implies that the eigenstate of $ H(\gamma_{ix},\gamma_{iy},k_z) $ is independent of $ \gamma $. As this eigenstate, independent of the reduced Brillouin zone (rBZ), spans the full Hilbert space in any distinct $ l $ which only lies in the occupied (or unoccupied) sector, and within this sector, the orbital basis is fixed in different $ (\gamma_i) $'s therefore, $ \Delta \mathscr{P}_{l,\gamma_1\gamma_2} = \mathscr{P}_l(\gamma_1) - \mathscr{P}_l(\gamma_2) $ should be an integer. Similarly, as different rotational representations cannot mix at $ \gamma_i $'s, the total polarization can be summed up from the polarization with different angular momentum sectors $ \mathscr{P}(\gamma_i) = \sum_{l\in \textrm {occupied}}\mathscr{P}_l(\gamma_i) $, implying $ \Delta \mathscr{P}_{\gamma_1\gamma_2} \in \mathbb{Z} $. This implies that integer number of charge is pumped in between any two rotation-symmetry invariant points in the rBZ. As the Chern number is zero in the $k_z$ plane, the charge pumped over one half-period will return back in other half of rBZ. Thus the phenomena is characterized as  \enquote{Returning Thouless Pump} (RTP) and the integer $\mathbb{Z}$ is called RTP invariant.


For the Hopf model as shown in fig (\ref{fig:Hopf_WCC}), we can see for stacking the layers in z-direction, $\Delta \mathscr{P}_{\Gamma X} = 1$ and $\Delta \mathscr{P}_{\Gamma M} = 1$ due to rotation symmetry along z axis in reduced Brillouin Zone. While for stacking layers in other direction HWF centres are confined and do not show multicellularity. For the Hopf-Chern lattice model it is elaborately addressed in the main part.

\subsection{\label{C2} $C=2$ Model phase diagram}
The momentum space Hamiltonian for $C=2$ phases is given by 
\begin{align}
H_{\textrm C2}(\textbf{k}) = d_0(\textbf{k})\mathbb{1} + d_x(\textbf{k})\sigma_x + d_y(\textbf{k})\sigma_y + d_z(\textbf{k})\sigma_z,
\label{Eq:C2}
\end{align}
where, by incorporating $t_a$ - $t_b = \delta$, the components can be expressed as,

\begin{align}
&d_0(\mathbf{k}) = (2 t_a - \delta)(\cos k_x+\cos k_y)\nonumber\\
&d_x(\mathbf{k})= 2 t_2 (\sin k_x \sin k_z +\cos k_z \sin k_y)\nonumber\\ 
&d_y(\mathbf{k}) = 2 t_2 (-\sin k_y\sin k_z +\cos k_z \sin k_x)\nonumber\\
&d_z(\mathbf{k}) =(h+\delta(\cos k_x+\cos k_y)) +4t'\cos k_x \cos k_y 
\end{align}
 The phase diagram for the parameters $t'=1$, $t_2=-2$ is illustrated below as a function of $h$ and $\delta$.
\begin{figure}[h!]
\centering
	\includegraphics[width=0.56\linewidth]{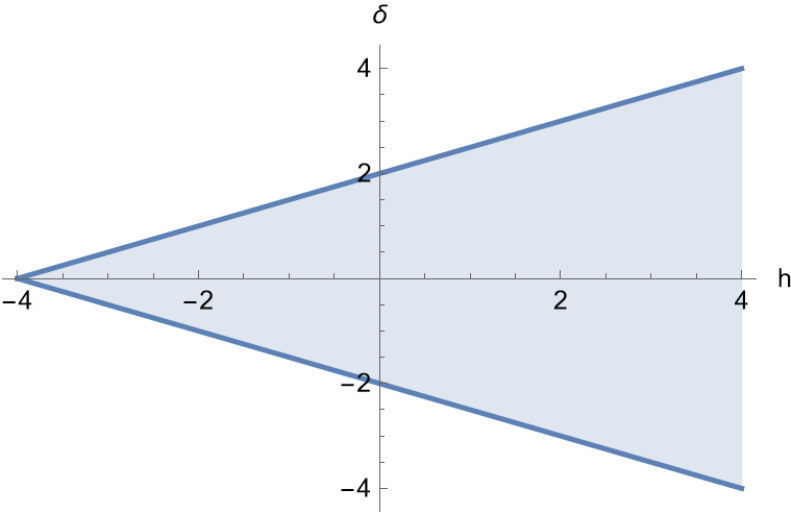}
	\caption{The shaded region in the diagram represents the phases where the Chern number $C=2$ for the parameters $t'=1$, $t_2=-2$.}
	\label{fig:A2}
\end{figure}

\section*{References:}
\bibliography{main.bib}

\end{document}